# A Framework for Creating a Distributed Rendering Environment on the Compute Clusters

Ali Sheharyar

IT Research Computing
Texas A&M University at Qatar
Doha, Qatar

Othmane Bouhali

Science Program and IT Research Computing
Texas A&M University at Qatar
Doha, Qatar

*Abstract*—This paper discusses the deployment of existing render farm manager in a typical compute cluster environment such as a university. Usually, both a render farm and a compute cluster use different queue managers and assume total control over the physical resources. But, taking out the physical resources from an existing compute cluster in a university-like environment whose primary use of the cluster is to run numerical simulations may not be possible. It can potentially reduce the overall resource utilization in a situation where compute tasks are more than rendering tasks. Moreover, it can increase the system administration cost. In this paper, a framework has been proposed that creates a dynamic distributed rendering environment on top of the compute clusters using existing render farm managers without requiring the physical separation of the resources.

## I. INTRODUCTION

### A. Background

Rendering is a process of generating one or more digital image(s) from a model or a collection of models, characterized as a virtual scene. A virtual scene is described in a scene file that contains the information such as geometry, textures, lights, etc. It is modelled in a 3D modelling application. Most commonly used modelling applications are Blender [1], Autodesk 3D Studio Max [2] and Autodesk Maya [3]. All modelling applications have a user interface with a drawing area where users can create a variety of geometrical objects, manipulate them in various ways, apply textures, and even animate etc. **Error! Reference source not found.** shows the user interface of Blender 3D modelling application [1]. A virtual scene is then given to the renderer that generates a set of high quality images later to be used to produce the final animation. Some of the most popular renderers are mental ray [4], V-Ray [5] and Blender [1].

Rendering is a compute-intensive and time-consuming process. Rendering time for an individual frame may vary from a few seconds to several hours. The rendering time depends on scene complexity, degree of realism (shadows, lights etc.) and output resolution. For example, a short animation project may be two-minutes in length, but at 30 frames per second (fps), it contains 3,600 frames. An average rendering time for a fairly simple frame can be approximately 2 minutes, resulting in a total of 120 hours. Fortunately, rendering is a highly parallelizable task as rendering of individual frames does not depend on any other frame. In order to reduce the total rendering time, rendering of individual frames can be distributed to a group of computers on the network. An animation studio, a company dedicated to production of animated films, typically has a cluster of computers dedicated to render the virtual scenes. This cluster of computers is called a render farm.

### B. Objectives

In a university environment, it can be complicated to do the rendering because many researchers do not have access to a dedicated machine for rendering [6]. They do not have access to a rendering machine for a long time as it may become unavailable for other research use. Moreover, they can lose their data due to hardware failure. By creating a distributed rendering environment, these problems can be addressed. However, some universities have one or more compute clusters that are normally used to perform high performance computing tasks. Distributed rendering on a compute cluster is possible, but it requires a lot of manual interaction with a cluster's job scheduler and is cumbersome.

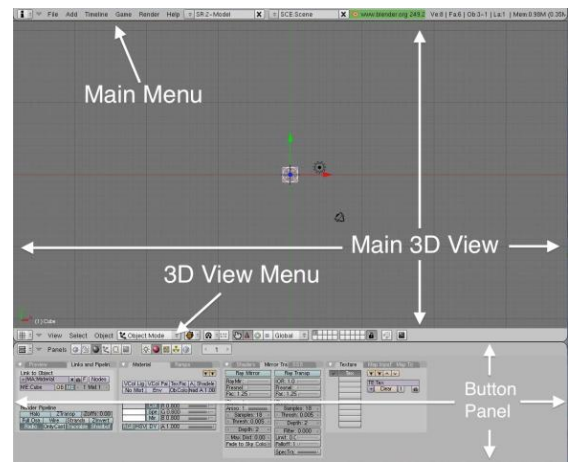

Fig.1.    Screen shot of the Blender user interface

In this paper, a framework has been presented to create a render farm-like environment on a compute cluster by using existing render farm software. This way, researchers and students will be presented with an interface similar to what typically animation studios have. Moreover, it does not require manual interaction with the cluster's job scheduler and makes the rendering workflow smoother.

*C. Related Work*

There has been some work on doing the image rendering on the cluster and on grids [7]. In this section, related work is briefly presented.

Huajin and Bing [8] discuss the design and implementation of a render farm manager based on OpenPBS. OpenPBS is a resource managers used for compute clusters. They have proposed to extend the OpenPBS functionality in order to facilitate the render job management. They maintain a state table to hold the render jobs status. They implement a new command "qsubr" to submit the job and another command "qstatr" to monitor the render jobs. They also provide a web-based interface in order to facilitate the job submission and monitoring.

Gooding et al. [6] talk about implementation of distributed rendering on diverse platforms rather than a single cluster. They consider utilizing all available resources such as recycled computers, community clusters and even TeraGrid [9]. One benefit of this approach is that it gives access to diverse computing resources, but on the other hand it requires significant changes in the infrastructure. They require adding a couple of new servers to host the software for job submission and distribution to render machines. They also need a new central storage system because it is essential for the network distribution of the render job's resources (textures etc.) so that all render nodes could access it and save the output back. It is obvious that it requires a change in networking infrastructure. They offer only command line interface for job submission and support, only RenderMan rendering engine [10].

Anthony et al. [11] propose a framework of distributed rendering over Grid by following two different approaches. One approach is to setup the portal through which a user can submit a rendering job on-demand and download the rendered images from the remote server at a later time. Another approach is to submit the job to the Grid through a GUI plug-in within the 3D modelling software where every rendered image will be sent back individually to the client concurrently. Both of these approaches require significant effort for implementation. They also talk about compression methods that are beyond the scope of this paper.

All of these approaches focus on implementing the render farm manager (or job manager). They provide users a way to interface to submit and control the render jobs in the form of command line using SSH, online web portal and/or plug-in within 3D modelling software. In summary, they need a significant amount of time and resources to implement all the nitty-gritties of various software components. In the next section, a new approach will be proposed.

This paper is divided in to several sections. Section II and III give an overview of render farms and compute clusters respectively. Section IV describes the current approaches and proposed approach to render the computer animation in distributed computing environments. Section V presents the experimental results and, finally, section VI and VII presents the conclusion and future work respectively.

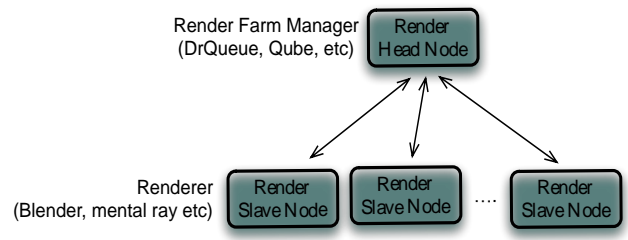

Fig.2.  Render farm architecture

## II. RENDER FARM

A render farm is a cluster of computers connected to a shared storage dedicated to render the compute-generated imagery. Usually, a render farm has one master (or head) machine (or node) and multiple slave machines. The head node runs the job scheduler that manages the allocation of resources on slave machines to jobs submitted by users (artists). **Error! Reference source not found.** shows the client/server architecture of a render farm. In the diagram, Render Head Node runs the server software of render farm manager, whereas, Render Slave Node runs the client software.

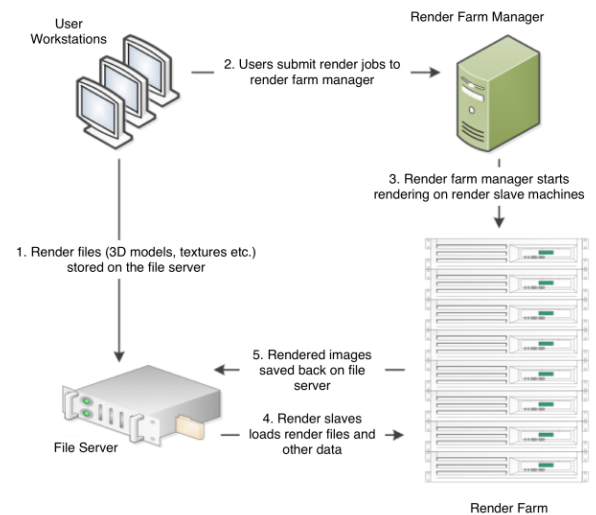

Fig.3.  Rendering workflow

A typical rendering workflow (see **Error! Reference source not found.**) can be described in the following steps:

*1) Artists create the virtual scenes on their workstations.*

*2) Artists store their virtual scene files, textures etc. on the shared storage.*

*3) Artists submit rendering jobs to the queue manager, a software package running on the head node.*

*4) Queue manager divides the job into independent tasks and distributes them to slave machines. A task could be rendering of one full image, a few images, or even a sub-section (tile) of an image. A job may have to wait in the queue depending on its priority and load on the render farm.*

*5) Slave machines read the virtual scene and associated data from the shared storage.*

*6) Slaves render the virtual scene and save the resulting image(s) back on the file server.*

*7)    User is notified of job completion or errors, if any.*

A render queue manager (also known as render farm manager or job scheduler), typically a client-server package facilitates the communication between the slaves and master. The head node runs the server component whereas all slave nodes run the client component of the render queue manager; although some queue managers have no central manager. Some common features of queue managers are prioritization of queue, management of software licenses, and algorithms to optimize the throughput in the farm. Software licensing handled by the queue manager might involve dynamic allocation of available CPUs or even cores within CPUs.

### III.    COMPUTE CLUSTER

A compute cluster is a group of computers linked with each other through a fast local area network. Clusters are used mainly for computational purposes rather than handling the IO-oriented operations such as databases or web services. For instance, a cluster might support weather forecast simulation or flow dynamics simulation of a plane wing.

A typical compute cluster comprises one head node and multiple compute (or slave) nodes. All clusters run a resource management software package that accepts jobs from users. They preserve them until they are run, run the jobs, and deliver the output back to user. **Error! Reference source not found.** shows the architecture of a compute cluster. Compute Head Node runs the server software of the resource manager, whereas, Compute Slave Node runs client software.

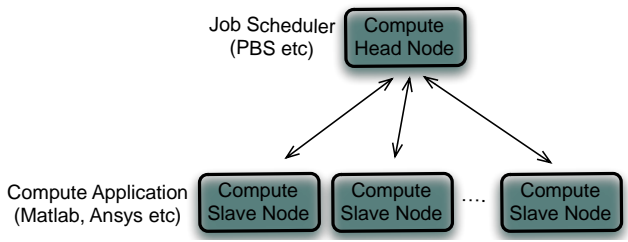

Fig.4.    Compute cluster architecture

A typical workflow to execute a job on a cluster is described below:

*1)    User prepares the job file that contains some parameters and path to the executable or script to run. For instance, the amount of memory and number of CPU cores required are specified.*

*2)    User submits the job file to the job scheduler. Submission is done usually over SSH terminal but some job schedulers also offer the web interface for job submission.*

*3)    Scheduler puts the job in to appropriate queue.*

*4)    When the job's turn comes, scheduler allocates the resources and starts the execution of executable/script specified in job description on the allocated slave node.*

*5)    When job finishes its execution, the output and error log is saved to disk. The scheduler can terminate a job if its execution time exceeds a predefined amount of time.*

*6)    User is notified of job completion.*

### IV.    RENDERING ON A COMPUTE CLUSTER

It is clear that both render farms and compute clusters have similar architecture. Both of them contain one master machine (or head node) and one or more slave machines. Both run a resource manager software package and both have similar workflows. A render farm can be considered as a special kind of compute cluster that uses resource manager and other software (renderers) specific to rendering the computer-generated imagery.

This section focuses on running the render-farm ecosystem over a compute cluster. First, current approaches to solve this problem and their limitations have been described. Then, a new approach has been proposed that not only can present existing and familiar interfaces to users but also requires less implementation effort.

#### A.    Proposed Cluster-based Rendering Framework

As mentioned above, all of existing work [6][8][11] have focused on developing all components of rendering job management and scheduling from scratch. However, this paper proposes a new approach using existing open-source or commercial render farm managers, meeting the requirements mentioned later, and using the compute cluster's resource with minimal or no change in existing setup.

Recall from earlier sections that a render farm manager has client/server architecture. The server software (r-server) runs on the head node whereas the client component (r-client) runs on slave nodes. The key difference between existing approaches and the proposed approach is that current approaches schedule the render jobs submitted by users directly to the cluster or grids and manage their state and execution process themselves. However, the new approach proposes to schedule the client-component of the render farm manager rather than the render jobs directly. The server component can either run on the cluster's head node (compute head node) along with cluster's existing job manager or a new server machine (render head node) that can be added to the existing environment.    The render head node also runs a software module called Rendering on Cluster Meta Scheduler (RCMS) (see **Error! Reference source not found.**).

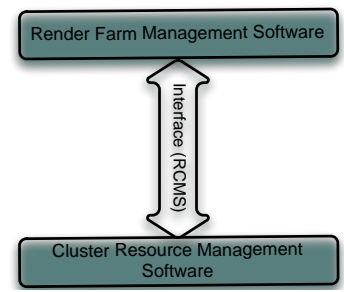

Fig.5.    Interface between render farm manager and cluster resource manager

The RCMS scheduling module queries the render farm manager and dynamically adjusts the number of active r-client jobs. Each r-client job appears to the r-server as a render slave node. The number of active r-client jobs depends on the current load of the compute cluster and the number of render jobs in queue. If there are pending render jobs, RCMS can submit new r-client jobs to the compute cluster. It also kills the active r-client jobs if there is no render job in queue and releases the resources to make them available for compute jobs. It maintains a state table to keep track of r-client jobs submitted to the cluster (see **Error! Reference source not found.**).

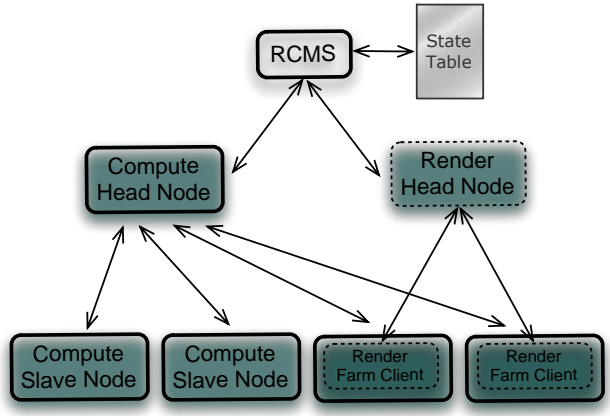

The RCMS module should be able to talk to render farm manager and cluster resource manager in order to keep the rendering system running effectively. Interfacing with cluster managers is trivial because most of cluster managers support at least command line interface for job submission etc. On the other hand, not all render farm managers are compatible with this framework. There is a set of requirements that a render farm has to meet in order to be compatible with the proposed framework. These requirements are described in the following sub-section.

Fig.6.        Render farm integration with the compute cluster

### B.  Render Farm Manager Compatibility Requirements

There are several open-source and commercial render farm managers available. But, in order to be compatible with proposed distributed rendering on cluster framework, it should meet the following requirements:

*1)    Job control API: Render farm manager should support a set of calls to query the information about the render jobs submitted by users. This interface can be either command-line programs or API of any programming language such as Python, Ruby, C/C++ etc.*

*2)    Failsafe rendering: The render farm manager should be able to detect failed or incomplete rendered images. As an r-slave job can run only for a fixed amount of time. After that, the cluster job manager will terminate it. It is important that the render farm manager should detect the incomplete rendered images and reschedule them.*

*3)    Automatic client recognition: The server should automatically detect active clients on slave nodes. Clients should send so-called heartbeats to the server, so that the server will automatically know their existence. It is required as clients are expected to be active dynamically over a set of compute nodes in the cluster.*

*4)    Supervisor required: Some render farm managers do not need the server component to manage the resources. However, the proposed approach requires that render farm management software has a supervisor to centrally control the jobs and resources.*

**Error! Reference source not found.** shows some of the popular render farm managers along with some features. As it is clear that Smedge and Spider are not compatible with the proposed framework because they do not support either supervisor and/or job control API.

Cluster Resource Manager Compatibility Requirements

All major cluster resource managers like PBS and LSF support at least SSH over command-line interface for user interaction. Some resource managers also support online web-interface.

Proposed distributed rendering framework requires that the cluster resource manager should have support for the following operations via command-line:

*1)    Job submission: A cluster manager should support job submission via command line. RCMS will prepare a job file that will specify the desired resources like number of cores, memory and execution time.*

*2)    Query jobs: It should support querying the currently active jobs by their names. RCMS will use its own naming scheme in order to identify the r-client jobs.*

*3)    Job deletion: As r-client jobs on cluster will be*

TABLE I.  RENDER QUEUE MANAGEMENT SOFTWARE

| Name | Supported 3D Applications | Supervisor Required? | Job Control API |
|---|---|---|---|
| DrQueue | Blender, Maya, mental ray, Pixie, command-line tools | Yes | Yes |
| Qube! | Maya, mental ray, SoftImage, RenderMan, Shake, command-line tools | Yes | Yes |
| Smedge | 3ds max, After Effects, Maya, mental ray, SoftImage | No | Yes |
| Spider | Maya | No | No |
| RenderPal | 3ds max, Blender, Cinema 4D, Houdini, Maya, mental ray, SoftImage | Yes | Yes |
| ButterflyNetRender (BNR) | All major applications and command-line tools | Yes | Yes |

dynamically deleted in order to release the cluster resources to be used for other computation tasks, it is necessary that cluster resource manager support this feature.

TABLE II.    HARDWARE AND SOFTWARE SPECIFICATION OF TEST PLATFORMS

| Machine | Dell 690 | Suqoor (single node) | HP Z800 |
|---|---|---|---|
| CPU | 2x Intel Clovertown X5355 @ 2.66 Ghz | 2x Intel Harpertown E5462 @ 2.80 GHz | 2x Intel Westmere-EP X5650 @ 2.66 GHz |
| Cores (per CPU) | 4 | 4 | 6 |
| Threads (per CPU) | 4 | 4 | 12 |
| L2 Cache  (per CPU) | 8 MB | 12 MB | 12 MB |
| CPU Launch Date | Q4'06 | Q4'07 | Q1'10 |
| Memory | 16 GB | 16 GB | 16 GB |
| Operating System(s) | Win XP 64-bit/Cent OS 6 (64-bit) | SuSE Linux Enterprise Server 10 (64-bit) | Red Hat Enterprise Linux 5 (64-bit) |
| GPU | Quadro FX 4600 | None | Quadro Plex 6000 |

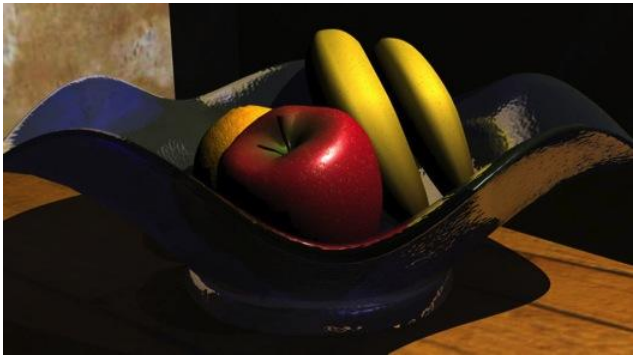

Fig. 7.    A 3D virtual scene modeled in Maya

## V.    EXPERIMENTAL RESULTS

As a proof of concept, a prototype of proposed distributed rendering framework is implemented and benchmarked. In this section, the benchmark results are presented. The prototype uses PipelineFX Qube [12] as render farm manager and PBS [13] for cluster resource management. It is implemented in Python language. The compute cluster (named *Suqoor*) at Texas A&M University at Qatar [14] has been used as a test environment. Out of ten available licenses for Qube, one is consumed by Qube Supervisor that manages all render jobs and remaining nine licenses are used by Qube workers. Each worker requires one license. A virtual scene (see **Error! Reference source not found.**) that comes with Autodesk Maya [3], a 3D modelling application, is used for the benchmarking. This scene has a 30-second long animation that comprises of 720 frames at 24 frames per second. For performance comparison, the same animation has been rendered (software rendering) on *Suqoor* and three other workstations as well. Software rendering refers to a rendering process that is unassisted by any specialized graphics hardware (such as graphics processing units or GPUs). Hardware rendering, utilizing the graphics hardware for rendering, cannot be performed on *Suqoor* due to lack of graphics hardware. However, performance of hardware rendering on workstations have also been compared to software rendering on *Suqoor*.

**Error! Reference source not found.** summarizes the hardware specification of the workstations and single compute node of the compute cluster (*Suqoor*). Note that two of the workstations have the same hardware specification (Dell 690) but have different operating systems. One has Windows XP x64 and other has Cent OS 6.

**Error! Reference source not found.** shows the average rendering time per frame on a single node (8 cores) of *Suqoor* and other workstations. Note that rendering time on a single compute node of cluster having 8 cores (25.22s) and HP Z800 workstation having 12 cores (25.86s) differs just by a fraction of a second. It has also been observed that rendering on Windows XP is almost 2.76 times slower than CentOS Linux on the same hardware configuration.

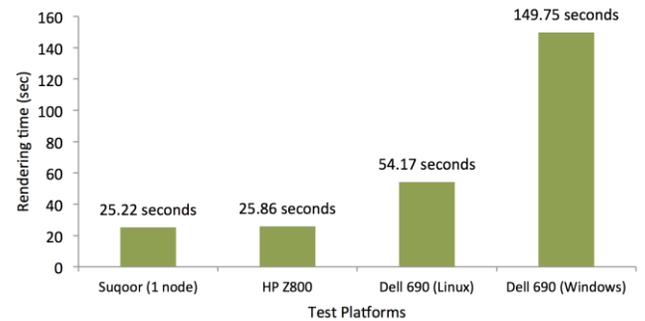

Fig. 8.    Average rendering time per frame (software rendering)

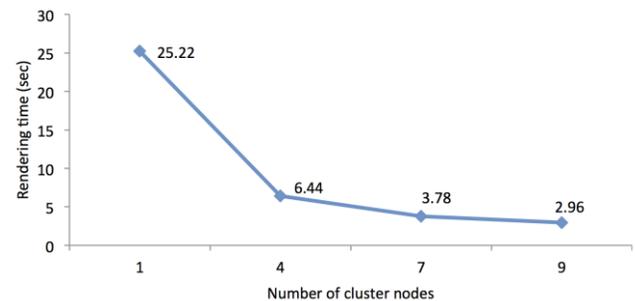

Fig. 9.    Average rendering time per frame versus number of cluster nodes (software rendering)

**Error! Reference source not found.** shows the average rendering time per frame by using 1, 4, 7 and 9 compute nodes, respectively. Due to the limited number of available Qube

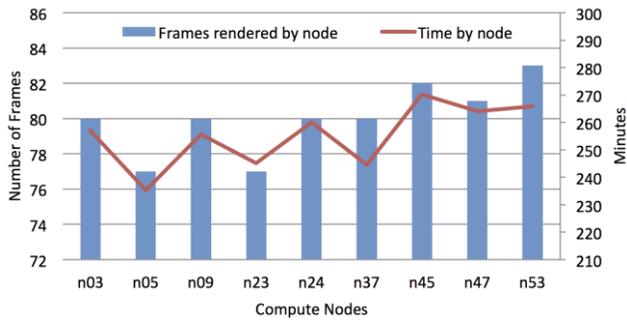

Fig. 10.    Number of frames rendered and accumulated time spent

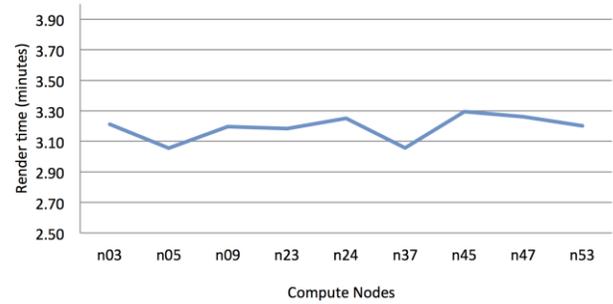

Fig. 11.    Average rendering times per frame with respect to compute nodes

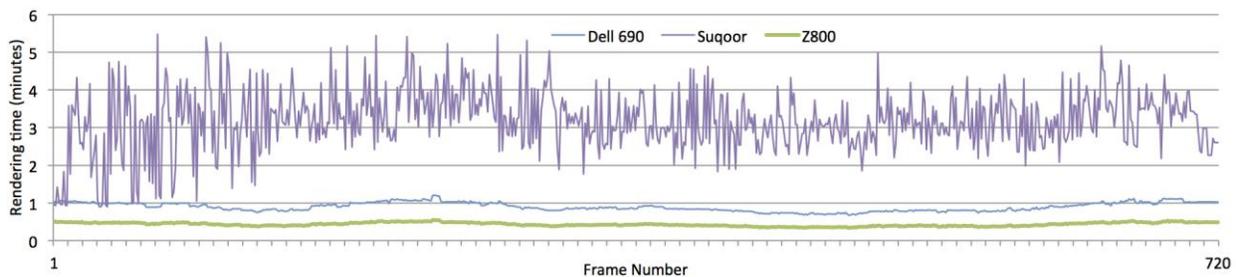

Fig. 12.    Rendering time of individual frames (software rendering)

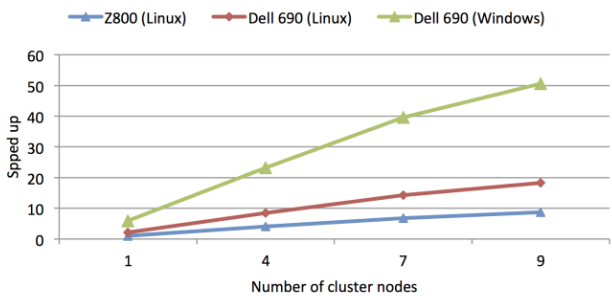

Fig. 13.    Speed up with cluster (software rendering)

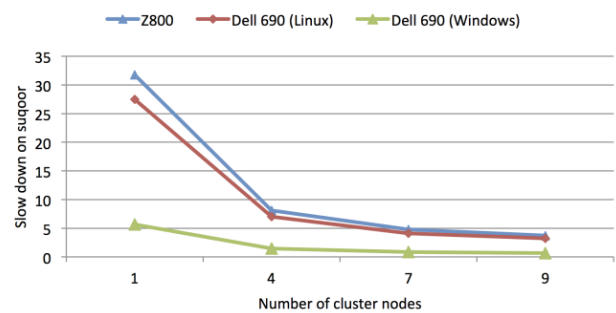

Fig. 14.    Software rendering on cluster versus hardware rendering on Windows and Linux workstations

licenses (10), the experiment could not be performed with more than 9 compute nodes.

**Error! Reference source not found.** shows the number of rendered frames and aggregated rendering time spent by each compute node. It shows how the Qube supervisor has distributed the render jobs across compute nodes. With an average distribution of 80 frames per node, it is apparent that work distribution is almost equal.

**Error! Reference source not found.** shows the average rendering times per frame with respect to each compute node. The variation in the result can be characterized to the variation in the 3D model complexity from different view angles.

**Error! Reference source not found.** shows the rendering time of all frames. Remember that there are 720 frames in the animation. It is observed that workstations render the frames one after the other by utilizing multiple CPU cores for single-frame rendering. On the other hand, *Suqoor* renders multiple frames on individual compute nodes. The numbers of active frames depend on the number of available cores on the compute nodes by assigning each core to an individual frame. Due to this, rendering time of individual frames is higher on *Suqoor* than other workstations.

**Error! Reference source not found.** shows the performance speed-up with respect to other platforms. For instance, a cluster with nine compute nodes performs nearly 51 times better than the Dell 690 workstation with Windows, nearly 18 times better than the Dell 690 workstation with Linux, and nearly 9 times better than the HP Z800 workstation.

**Error! Reference source not found.** compares the performance of software rendering on *Suqoor* to hardware rendering on other workstations. This plot shows how fast hardware rendering is on other workstations with respect to *Suqoor*. Remember that *Suqoor* does not support hardware rendering. It is observed that hardware rendering, especially on Linux workstations, is remarkably faster than software. The performance gap is drastically reduced by using more nodes on the cluster.

## VI. CONCLUSION

This paper has presented a framework that creates a distributed rendering environment on a general-purpose compute cluster by using an existing render farm management application. It can be used to create the rendering environment similar to that of an animation studio in a university environment where users do not have exclusive access to the computers to perform time-consuming image renderings. The prototype of the proposed framework, using Qube! for render farm management and PBS for compute cluster management, has been implemented. The experimental results show that the compute cluster reduces the rendering time significantly in case of software rendering. Moreover, by using the existing render farm manager, the overall rendering workflow becomes efficient.

## VII. FUTURE WORK

One thing, where compute cluster lacks behind is the hardware rendering that is due to the absence of GPUs in the compute nodes. Texas A&M University at Qatar is soon expected to acquire a larger cluster that will also have GPUs in several compute nodes. For the future work, the same experiment will be repeated on the new cluster and performance of the hardware rendering will be analyzed. The new cluster is expected to outperform the workstation by a large margin.

### References

[1]   Blender, http://www.blender.org

[2]   Autodesk 3D Studio Max, http://www.autodesk.com

[3]   Autodesk Maya, http://www.autodesk.com

[4]   Mentay ray, http://www.mentalimages.com/products/mental-ray.html

[5]   V-Ray, http://chaosgroup.com/en/2/index.html

[6]   Gooding, S. Lee, Laura Arns, Preston Smith, and Jenett Tillotson. "Implementation of a distributed rendering environment for the TeraGrid." In Challenges of Large Applications in Distributed Environments, 2006 IEEE, pp. 13-21. IEEE, 2006.

[7]   Grid Computing, http://en.wikipedia.org/wiki/Grid_computing

[8]   Jing, Huajun, and Bin Gong. "The design and implementation of Render Farm Manager based on OpenPBS." In Computer-Aided Industrial Design and Conceptual Design, 2008. CAID/CD 2008. 9th International Conference on, pp. 1056-1059. IEEE, 2008.

[9]   TeraGrid, http://www.teragrid.org

[10]  RenderMan, http://renderman.pixar.com

[11]  Chong, Anthony, Alexei Sourin, and Konstantin Levinski. "Grid-based computer animation rendering." In Proceedings of the 4th international conference on Computer graphics and interactive techniques in Australasia and Southeast Asia, pp. 39-47. ACM, 2006.

[12]  Pipeline FX Qube!, http://www.pipelinefx.com

[13]  PBS Guide, http://hpc.sissa.it/pbs

[14]  High Performance Computing at Texas A&M University at Qatar, http://technology.qatar.tamu.edu/rc/2000.aspx